# Observation of a Knotted Electron Diffusion Region in Earth's Magnetotail Reconnection

Xinmin Li[1, *], Chuanfei Dong[1, *], Hantao Ji[2,3], Chi Zhang[1], Liang Wang[1], Barbara Giles[4], Hongyang Zhou[1], Rui Chen[5], and Yi Qi[6]

[1]*Center for Space Physics and Department of Astronomy, Boston University, Boston, MA, USA*

[2]*Department of Astrophysical Sciences, Princeton University, Princeton, NJ, USA*

[3]*Princeton Plasma Physics Laboratory, Princeton University, Princeton, NJ, USA*

[4]*NASA, Goddard Space Flight Center, Greenbelt, MD, USA*

[5]*Department of Physics, Auburn University, Auburn, AL, USA*

[6]*Laboratory for Atmospheric and Space Physics, Boulder, CO, USA*

[*]Correspondence to: X. Li (xli8@bu.edu) and C. Dong (dcfy@bu.edu)

**Key Points**

1) A knotted EDR was observed in Earth's magnetotail, with its reconnection plane tilted ~38° from the IDR

2) Significant variations in the guide field and Hall magnetic field structures are identified between the EDR and IDR

3) The results highlight the importance of 3D effects and suggest a complex multiscale coupling between the EDR and IDR

**Abstract**

Magnetic reconnection is a fundamental plasma process that alters the magnetic field topology and releases magnetic energy. Most numerical simulations and spacecraft observations assume a two-dimensional diffusion region, with the electron diffusion region (EDR) embedded in the same plane as the ion diffusion region (IDR) and a uniform guide field throughout. Using observations from Magnetospheric Multiscale (MMS) mission, we report a non-coplanar, knotted EDR in Earth's magnetotail current sheet. The reconnection plane of the knotted EDR deviates by approximately 38° from that of the IDR, with the guide field exhibiting both a 38° directional shift and a twofold increase in amplitude. Moreover, the Hall magnetic field is bipolar in the EDR but quadrupolar in the IDR, indicating different Hall current structures at electron and ion scales. These observations highlight the importance of three-dimensional effects and illustrate the complexity of multiscale coupling between the EDR and IDR during reconnection studies.

**Plain Language Summary**

Magnetic reconnection is a key process in space where magnetic field lines break and reconnect, releasing energy and accelerating particles. Scientists often use two-dimensional (2D) models to describe this process, assuming the entire region where reconnection occurs lies in the same plane. However, recent observations from NASA's Magnetospheric Multiscale (MMS) mission challenge this idea. In this study, we report a special case where the small-scale electron diffusion region (EDR) is not in the same plane as the larger-scale ion diffusion region (IDR). Instead, the EDR is tilted by about 38°, and this misalignment leads to strong changes in the magnetic guide field and Hall magnetic field structures. We refer to this tilted and non-coplanar structure as a "knotted EDR." These findings highlight the importance of three-dimensional effects in space plasmas and suggest that magnetic reconnection can behave in more complex ways than previously expected.

## 1. Introduction

Magnetic reconnection is a fundamental plasma process that enables the rearrangement of magnetic field topology and the conversion of magnetic energy into plasma kinetic and thermal energy [*Hesse and Cassak*, 2020; *Ji et al.*, 2022; *Q Lu et al.*, 2022]. It has been observed in a wide range of space and astrophysical environments, such as the solar atmosphere, solar wind, Earth's magnetosphere, and other planetary magnetospheres, playing a crucial role in particle acceleration, mass and energy transport, and planetary atmospheric escape [*V. Angelopoulos et al.*, 2008; *Dong et al.*, 2019; *Dong et al.*, 2018; *Dong et al.*, 2022; *Lin et al.*, 2003; *Oieroset et al.*, 2002; *T Phan et al.*, 2006; *Christopher T Russell and Elphic*, 1979; *R Wang et al.*, 2024; *R Wang et al.*, 2023; *Zhang et al.*, 2012]. Due to observation and computing resource limitations, magnetic reconnection is often described in a two-dimensional (2D) framework, assuming a well-defined reconnection plane and uniform physical quantities in the out-of-reconnection-plane direction. In collisionless plasma environments, the 2D reconnection model suggests that the reconnection diffusion region consists of two co-planar areas: an ion diffusion region (IDR) and an embedded electron diffusion region (EDR) [*Birn and Hesse*, 2001; *Fu et al.*, 2006; *Hesse et al.*, 1999; *Q Lu et al.*, 2010; *L Wang et al.*, 2015; *Yang et al.*, 2023], which has been confirmed by multiple spacecraft observations [*Burch et al.*, 2016; *Eastwood et al.*, 2010; *X M Li et al.*, 2019; *Øieroset et al.*, 2001; *Torbert et al.*, 2018; *Zhou et al.*, 2019].

The 2D framework has been extensively adopted in both numerical simulations and spacecraft observations, successfully capturing many key features of reconnection. However, some reconnection cases cannot be fully understood within the 2D framework. For instance, the reconnecting current sheet can locally deform by the tearing mode instability or lower-hybrid drift instability, leading to localized bending or twisting of the reconnecting current sheet [*Ahmadi et*

*al.*, 2025; *Cozzani et al.*, 2021; *R Ergun et al.*, 2019a; *Y-H Liu et al.*, 2013; *Y H Liu et al.*, 2018; *Yoon et al.*, 2002]. Several observations of magnetotail reconnection suggest that the X-line orientation can deviate significantly from the expected perpendicular alignment with the reconnection plane [*Pathak et al.*, 2022; *Qi et al.*, 2023], indicating the presence of 3D reconnection effects. Moreover, as turbulence develops within the reconnecting current sheet, the reconnecting current sheet becomes populated with 3D magnetic flux ropes and filamentary currents [*Che et al.*, 2011; *Che and Zank*, 2020; *Daughton et al.*, 2011; *X Li et al.*, 2022a; *X Li et al.*, 2022b; *S Lu et al.*, 2023; *R Wang et al.*, 2023; *R S Wang et al.*, 2016; *S Wang et al.*, 2020b], which again are deviations from the 2D model. These deviations confirm that 3D effects play a significant role. However, a clear understanding of how 3D effects influence the structure and dynamics of the EDR remains elusive.

In this work, we report a knotted EDR observed by MMS in Earth's magnetotail current sheet. Contrary to 2D assumptions, the observed knotted EDR is not co-planar with the IDR, exhibiting a significant angular deviation (38°) between the reconnection plane of the EDR and that of the larger-scale IDR. This results in a large difference in the guide field and Hall magnetic fields between IDR and EDR. The observations demonstrate the importance of 3D effects in understanding the structure and dynamics of the EDR.

## 2. Instrumentation

The data used in this work are obtained from three instruments onboard MMS. The Fluxgate Magnetometer (FGM) provides magnetic field measurements at a 128 Hz time resolution [*C. T. Russell et al.*, 2016]. The Electric Field Double Probes (EDP) measure the 3D electric field with a time resolution of 8,192 Hz [*R. E. Ergun et al.*, 2016; *Lindqvist et al.*, 2016]. Plasma moments and distribution data are acquired from the Fast Plasma Investigation (FPI), with a time resolution of 30 (150) ms for electrons (ions) [*Pollock et al.*, 2016].

## 3. Observation and Analysis

### 3.1 Overview of the Ion Diffusion Region

During 12:05-12:22 UT on 11 July 2022, MMS was located at [-15.0, -4.0, 1.8] $R_E$ (where $R_E$ is Earth's radius, about 6,400 km) in the Geocentric Solar Ecliptic (GSE) coordinate system, within the plasma sheet of the magnetotail. During this interval, MMS observed an active magnetic reconnection event, previously reported by *Ou et al.* [2025]. Figure 1 presents an overview of the event as observed by MMS1 in the local current coordinate system (LMN). The LMN coordinate system is nearly consistent with GSE coordinate, with the transformation matrix from GSE to LMN given by **L** = [0.9985, 0.0535, 0.0057], **M** = [-0.0507, 0.8987, 0.4357], and **N** = [0.0182, -0.4353, 0.9000]. Here, **N** is the maximum variance from the minimum variance analysis (MVA) based on the electric field data from 12:12:30 to 12:14:30 UT [*Sonnerup and Scheible*, 1998]. **M** = **N** × **L**$_0$, where **L**$_0$ is the maximum variance from the MVA based on the ion bulk flow velocity. And **L** completes the right-handed system. The MVA method is not applied to the magnetic field because the spacecraft did not directly cross the current sheet.

The ion flow **V**$_{i, L}$ (blue trace in Figure 1d) reverses from tailward ($V_{i,L} < 0$) to earthward ($V_{i,L} > 0$) between 12:05 to 12:20 UT, with the reversal occurring at approximately 12:13:20 UT. Accompanying this reversal, the magnetic field **B**$_N$ (red trace in Figure 1c) changes from negative to positive, indicating that MMS encounters an ongoing reconnecting current sheet. The magnetic field **B**$_L$ (blue trace in Figure 1a) remains negative until 12:18 UT, when MMS is inside the earthward outflow region. The transition of **B**$_L$ suggests that MMS is located on the Southern side of the reconnecting current sheet prior to 12:18 UT, when MMS crosses the neutral plane of the current sheet and enters the northern side.

With a guide field $B_g = 5$ nT, $\mathbf{B}_M$ (green trace in Figure 1c) exhibits a positive-negative-positive variation, indicating that MMS sequentially traverses three of the four quadrants of the Hall magnetic field structures: the southern-tailward quadrant, the southern-earthward quadrant, and the northern-earthward quadrant. In addition, inside the tailward outflow region, a magnetic flux rope is observed around 12:11 UT, characterized by the bipolar variation in $\mathbf{B}_N$ and the enhancement in the magnitude of the magnetic field (black trace in Figure 1c). These observations suggest that MMS crosses the IDR of an active magnetotail reconnection event.

Near the reversal point of $\mathbf{V}_{i,L}$ and $\mathbf{B}_N$, $\mathbf{B}_L$ remains negative but with a small magnitude, indicating that MMS is located close to the X-line of the reconnecting current sheet. Based on the collisionless reconnection model, the EDR, corresponding to an electron-scale current layer, is expected to form in the vicinity of the X-line [*Burch et al.*, 2016; *Hesse et al.*, 1999; *Torbert et al.*, 2018]. In this event, a current layer (Figure 1h) is indeed observed around the X-line (~12:13:20 UT). Inside this current layer, high-speed electron jets are detected in both the L and M directions (Figure 1g), with peak values of $\mathbf{V}_{e,L}$ and $\mathbf{V}_{e,M}$ exceeding 2,500 km/s. The energy dissipation rate $\mathbf{J} \cdot (\mathbf{E} + \mathbf{V_e} \times \mathbf{B})$ [*Zenitani et al.*, 2011], along with its parallel and perpendicular components, is shown in Figure 1i. A positive dissipation rate, dominated by the parallel components, is observed inside the current layer, consistent with the previously reported signatures of EDRs in magnetotail current sheets [*X M Li et al.*, 2019; *Torbert et al.*, 2018; *R S Wang et al.*, 2020a]. These observations indicate that MMS encountered the EDR associated with magnetotail reconnection (Additional supporting evidence for this assertion is presented in the next section).

In contrast to a typical EDR, where the current layer is primarily supported by the variation of magnetic field component $\mathbf{B}_L$, the current layer in this observed EDR is supported by the variations in both $\mathbf{B}_L$ and $\mathbf{B}_M$. Notably, a distinct bipolar structure is observed in $\mathbf{E}_N$ within the EDR (red trace in Figure 1f), deviating from the expected positive Hall electric field $\mathbf{E}_N$ typically seen on the southern side of the reconnecting current sheet. These deviations from the standard 2D reconnection configuration suggest that the structure of this EDR cannot be fully described in the LMN coordinate system defined for the IDR. To better characterize the local structure of this EDR, a new coordinate system is introduced and discussed in the following section.

### 3.2 Knotted Electron Diffusion Region

Figure 2 presents a zoomed-in view of the EDR in a newly defined local current coordinate system (LMN'), which is established in the vicinity of the EDR using a hybrid method. The $\mathbf{M}'$ direction is aligned with the maximum current density observed inside the EDR at 12:13:18.432 UT. The $\mathbf{N}'$ direction is defined as $\mathbf{N}' = \mathbf{L}'_0 \times \mathbf{M}'$, where $\mathbf{L}'_0$ corresponds to the direction of maximum variance obtained from a minimum variance analysis of the magnetic field data between 12:13:17 UT and 12:13:20 UT [*Sonnerup and Scheible*, 1998]. The L' direction is chosen to complete the right-hand system. The transformation matrix from the original LMN coordinate system (used for the IDR and shown in Figure 1) to the LMN' system (used for the EDR) is given by $\mathbf{L}'$= [0.7826, 0.5865, -0.2085], $\mathbf{M}'$= [-0.6065, 0.7939, -0.0437] and $\mathbf{N}'$ = [0.1399, 0.1606, 0.9770].

The relative orientation between the LMN and LMN' coordinate systems is shown in the upper-left panel of Figure 3, where the blue and yellow lines represent the LMN and LMN' systems, respectively. The LMN' coordinate system is derived from the LMN system via a counterclockwise rotation of approximately 38° about an axis close to N. The resulting local reconnecting magnetic field direction within the EDR, L', lies between the global L and M directions, while the guide field direction M' forms an angle of ~38° with the original M direction. The normal direction N' remains approximately aligned with the current sheet normal. This significant deviation from the global reconnection plane highlights a localized, non-coplanar geometry of the EDR, which we refer to as a knotted EDR (as illustrated in Figure 3).

Within the knotted EDR, the plasma number density remains nearly constant (Figure 2a). The reconnecting magnetic field $\mathbf{B}_{L'}$ (blue trace in Figure 2b) changes from +5 nT to -5 nT, with a reversal point occurring at 12:13:18.4 UT. Around the reversal point, the magnetic field $\mathbf{B}_{N'}$ exhibits a positive enhancement, suggesting that MMS crosses the EDR on the +**L**' side of the X-line (illustrated in the lower-left panel of Figure 3). The magnetic field $\mathbf{B}_{M'}$ has a background value of 10.5 nT, about twice as large as the reconnecting magnetic field $\mathbf{B}_{L'}$. Moreover, $\mathbf{B}_{M'}$ shows a positive enhancement near the center of the EDR, with a magnitude of about 3 nT. This enhancement will later be interpreted as the Hall magnetic field.

Figure 2c shows the electron bulk flow velocity. The electron flow $\mathbf{V}_{e,M'}$ is negative and reaches approximately 3500 km/s, which corresponds to about $0.7V_{Ae}$, where $V_{Ae}$ is the electron Alfven speed, calculated using $\mathbf{B}_{L'}$ =5 nT and n=0.85 cm$^{-3}$. This high-speed electron flow in the **M**' direction leads to an intense current density in the +**M**' direction (Figure 2d), as supported by the variation of the magnetic field $\mathbf{B}_{L'}$. In addition, two oppositely directed electron jets in $\mathbf{V}_{e,L'}$ (blue trace in Figure 2c) are observed within the EDR, corresponding to the electron inflow (negative jet) and outflow (positive jet), respectively. The electron jets contribute to the bidirectional Hall currents along the **L**' direction (Figure 2d), resulting in an enhancement of the magnetic field $\mathbf{B}_{M'}$, which constitutes the Hall magnetic field. Due to the presence of the strong guide field, only a unipolar Hall magnetic field is observed, consistent with previous simulations and observations [*Huba*, 2005; *X Li et al.*, 2023; *T D Phan et al.*, 2018; *Pritchett and Coroniti*, 2004; *S Wang et al.*, 2021].

The electric field $\mathbf{E}_{N'}$ (Figure 2e) exhibits a bipolar structure inside the EDR. Unlike the previous studies, where $\mathbf{E}_{N'}$, corresponding to the Hall electric field, points toward the current sheet center, $\mathbf{E}_{N'}$ is directed outward in this case. Typically, the Hall electric field is balanced by $(\mathbf{J} \times \mathbf{B})/nq$, where $\mathbf{J}$ is current density, $\mathbf{B}$ is magnetic field, $n$ is plasma number density, and $q$ is the elementary charge. Inside the EDR, where the ion flow is weak, $(\mathbf{J} \times \mathbf{B})/nq$ is primarily balanced by the electron term $-(\mathbf{V}_e \times \mathbf{B})_N$, which decomposes into $\mathbf{V}_{e,M}\mathbf{B}_L$ and $-\mathbf{V}_{e,L}\mathbf{B}_M$. In anti-parallel reconnection, $\mathbf{V}_{e,M}\mathbf{B}_L$ dominates and points toward the current sheet center [*Birn and Hesse*, 2001; *Q Lu et al.*, 2010]. Here, however, the guide field $\mathbf{B}_{M'}$ is approximately twice as strong as the $\mathbf{B}_{L'}$, and $\mathbf{V}_{e,L'}$ is comparable to $\mathbf{V}_{e,M'}$. Thus, $-\mathbf{V}_{e,L}\mathbf{B}_M$ becomes significant and opposes $\mathbf{V}_{e,M}\mathbf{B}_L$, leading to a net outward $\mathbf{E}_{N'}$ (Figure 2f). Similar phenomena have also been observed in previous observational reconnecting current sheets with strong guide fields [*T D Phan et al.*, 2018]. Furthermore, inside the EDR, the electrons are decoupled from the magnetic field lines, leading to the generation of the non-ideal electric fields ($\mathbf{E} + \mathbf{V}_e \times \mathbf{B} \neq \mathbf{0}$) inside the EDR, as expected (Figures 2f and 2g).

Figure 4 shows the multiple spacecraft observations of the EDR. The magnetic field $\mathbf{B}_{L'}$, observed by four spacecraft, exhibits similar profiles but with a time-shifted variation (Figure 4b), indicating that MMS1, MMS4, and MMS2 crossed the EDR in rapid succession, followed by MMS3 about 0.5 s later. This sequence is consistent with the relative positions of the spacecraft along the $\mathbf{N}'$ direction (as shown in the lower-right panel of Figure 3). Using the time delays between the four spacecraft crossings and their relative positions, we determine the relative velocity between the current sheet and spacecraft by the timing method [*Schwartz*, 1998]. The velocity is 30 km/s along the direction [-0.2621, -0.1191, 0.9576] in the LMN' coordinate system. The fact that this direction is nearly aligned with **N'** confirms that the LMN' coordinate system is appropriate for analyzing the local EDR structure. Given that the duration of the EDR is about 2.0 s (12:13:17.4-12:13:19.4 UT), the thickness of this EDR is estimated to be about 60 km, or 10 $d_e$, where $d_e$ = 5.8 km is the electron internal length based on an electron number density of n = 0.85 $cm^{-3}$.

All spacecraft observes positive enhancements in the magnetic field $B_{N'}$ (Figure 4d), indicating that all of them cross the EDR in the +**L'** direction. Among them, MMS1, MMS2 and MMS4 are close in the **N'** direction and therefore cross the EDR nearly simultaneously. As a result, the differences among their observations can primarily be attributed to spatial variations. Specifically, MMS1 and MMS4 are located farther from the X-line than MMS2 along the **L'** direction, and accordingly, they observe similar and stronger $B_{N'}$ compared to MMS2. This is consistent with the expected distribution of the reconnected magnetic field, which increases with the distance from the X-line along the **L'** direction.

In contrast to the highly similar observations from MMS1 and MMS2, the observations from MMS3 exhibit significant differences, including a weak bipolar variation in $B_{M'}$ (Figure 4c) and highly asymmetric electron inflow and outflow (green trace in Figure 4e, where the inflow magnitude exceeds that of the outflow). Since MMS3 is located between MMS1 and MMS2 in both the **M'** and **L'** directions, these discrepancies are attributed to its greater separation in the **N'** direction. Positioned farther from MMS1 and MMS2 in **N'**, MMS3 encounters the current sheet at a later time. Thus, the pronounced differences in MMS3's measurements compared to the other spacecraft suggest that the current sheet may have undergone temporal evolution.

Following the EDR crossing, a magnetic hole, characterized by a local reduction in magnetic field magnitude (Figure 4a), was observed by both MMS2 and MMS3 around 12:13:21 UT (marked by the green shaded region in Figure 4). This magnetic hole is induced by an electron vortex, with MMS2 and MMS3 crossing different quadrants of the structure. The fact that only MMS2 and MMS3 observe the magnetic hole indicates that its spatial scale is comparable to the spacecraft separation, i.e., less than 21 km (approximately 4 $d_e$).

In contrast to the EDR, where the electron temperature increases mainly in the parallel direction, the perpendicular electron temperature enhances inside the magnetic hole (Figure 4h). Furthermore, Figure 4i shows the pitch angle distribution of the electrons with energies between 2 and 30 keV, with the black lines representing the local loss cone angle. Flux enhancements are observed around 90°, and just outside the loss cone, indicating that the electrons are trapped by this magnetic hole. These observations are similar to previous observations of magnetic holes [*Huang et al.*, 2022; *Xie et al.*, 2024; *Yao et al.*, 2018]. As this magnetic hole is not directly related to the main topic of this paper and has been previously reported [*Ou et al.*, 2025], it will not be discussed in detail here.

## 4. Discussion and Conclusions

This study presents a detailed analysis of a magnetic reconnection event observed by MMS in the Earth's magnetotail current sheet, initially reported by *Ou et al.* [2025]. In this event, MMS observed both the larger-scale IDR and smaller-scale EDR. The IDR is analyzed in a local current coordinate system (LMN), where **L** and **N** define the reconnection plane, and **M** is the out-of-reconnection-plane direction, aligned with the guide field. Inside the IDR, the guide field is 5 nT, or ~0.25 $\mathbf{B}_{L,0}$ (where $\mathbf{B}_{L,0} \approx 20$ nT), indicating a weak guide field reconnection current sheet. The positive-negative-positive variation of the out-of-reconnection-plane magnetic field $\mathbf{B}_M$ suggests the presence of the quadrupolar Hall magnetic field. These features are consistent with the standard 2D reconnection configuration, and well described by the LMN coordinate system. However, unlike previously reported reconnection cases, where the EDR lies within the same reconnection plane as the IDR, the EDR in this event cannot be well described by the same LMN coordinate system. In other words, the EDR is not co-planar with the IDR reconnection plane. Thus, a separate local current coordinate system (LMN') is established, specifically centered on the EDR.

The normal direction remains basically unchanged between the IDR and EDR. And the relative orientation between LMN and LMN' reveals a rotation around the N-axis, leading to a significant angular deviation (~38°) between the reconnection plane of the EDR and that of larger-scale IDR, such that the reconnection magnetic field ($\mathbf{B}_{L'}$) inside the EDR involves both $\mathbf{B}_L$ and $\mathbf{B}_M$. In addition, the guide field changes substantially between the IDR and EDR, with a ~38° directional shift and an approximate twofold increase in magnitude. We refer to this localized, non-coplanar EDR structure as a knotted EDR (as illustrated in Figure 3). Moreover, the minimum directional derivative (MDD) [*Shi et al.*, 2005] analysis reveals that the direction of minimum variation inside the EDR is misaligned with the out-of-reconnection-plane directions of both the EDR and IDR (see Supplementary Figure), suggesting a fully three-dimensional configuration of the diffusion region.

Previous studies have shown that the electromagnetic drift waves can be generated inside the EDR, leading to kinking of the current layer that propagates along the out-of-reconnection-plane direction (**M**) [*Cozzani et al.*, 2021; *R Ergun et al.*, 2019a; *Robert E Ergun et al.*, 2019b; *Price et al.*, 2016]. The kinking of the current layer changes the orientation of the local current sheet, causing the out-of-reconnection-plane direction (**M'**) to have a significant component in the **N** direction. This scenario, however, differs from our observations of the knotted EDR, where the **M'** has a significant component along **L** rather than **N**. Furthermore, the electromagnetic drift waves are typically associated with significant plasma density perturbations within the EDR [*Cozzani et al.*, 2021; *Robert E Ergun et al.*, 2019b], which are not observed in this event (Figure 2a).

Alternatively, 3D simulations have suggested that the oblique tearing mode instability can lead to tilting of the current layer [*Daughton et al.*, 2011; *Y-H Liu et al.*, 2013; *Y H Liu et al.*, 2018], resulting in that **M'** has a significant component in the **L** direction, consistent with our observations. Notably, the observed tilt angle in this event (~38°) exceeds those typically reported in simulation studies. This difference may be attributed to the relatively weak guide field in this case, as oblique tearing modes are expected to produce more strongly tilted current layers under weaker guide field conditions [*Y-H Liu et al.*, 2013]. The exact mechanism responsible for the formation of such a 3D and strongly tilted knotted EDR remains to be further investigated.

While previous studies have highlighted non-orthogonal X-line orientations as evidence of 3D reconnection [*Pathak et al.*, 2022; *Qi et al.*, 2023], our results reveal a distinct 3D feature: a clear non-coplanarity between the IDR and EDR, referred to as a knotted EDR. This structural decoupling suggests that electron- and ion-scale reconnection may proceed in geometrically different planes, highlighting a more intricate 3D nature of magnetic reconnection. Furthermore, our observations raise an important question regarding the coupling between the EDR and IDR during the reconnection process. The EDR and IDR are not aligned in the same reconnection plane, and both the direction and magnitude of the guide field vary significantly between them. In addition, the IDR exhibits a typical quadrupolar Hall magnetic field, while the EDR is characterized by a bipolar Hall magnetic field. How the Hall magnetic field and the associated Hall current transition from the bipolar structure in the EDR to the quadrupolar configuration in the IDR remains an open question and deserves further investigation.

In summary, this work presents a unique observation of a non-coplanar, knotted EDR embedded within a larger-scale, two-dimensional IDR. The 38° deviation between the reconnection planes of the EDR and IDR, together with the distinct changes in the Hall magnetic field and guide field, highlights the importance of three-dimensional effects and the complexity of coupling between the EDR and IDR in magnetic reconnection.

**Acknowledgments**

This work is supported by DOE grant DE-SC0024639, NSF Grant AGS-2438328, and Alfred P. Sloan Research Fellowship. We thank the entire MMS team and instrument principal investigators for providing and calibrating data.

**Open Research**

All the MMS data used in this work are available at the MMS data center (https://lasp.colorado.edu/mms/sdc/public/about/browse-wrapper/). The data have been loaded, analyzed, and plotted using the SPEDAS software (Version 6.0) [*V Angelopoulos et al.*, 2019], which can be downloaded via the Downloads and Installation page (http://spedas.org/blog/).

**Figures and Figure Captions**

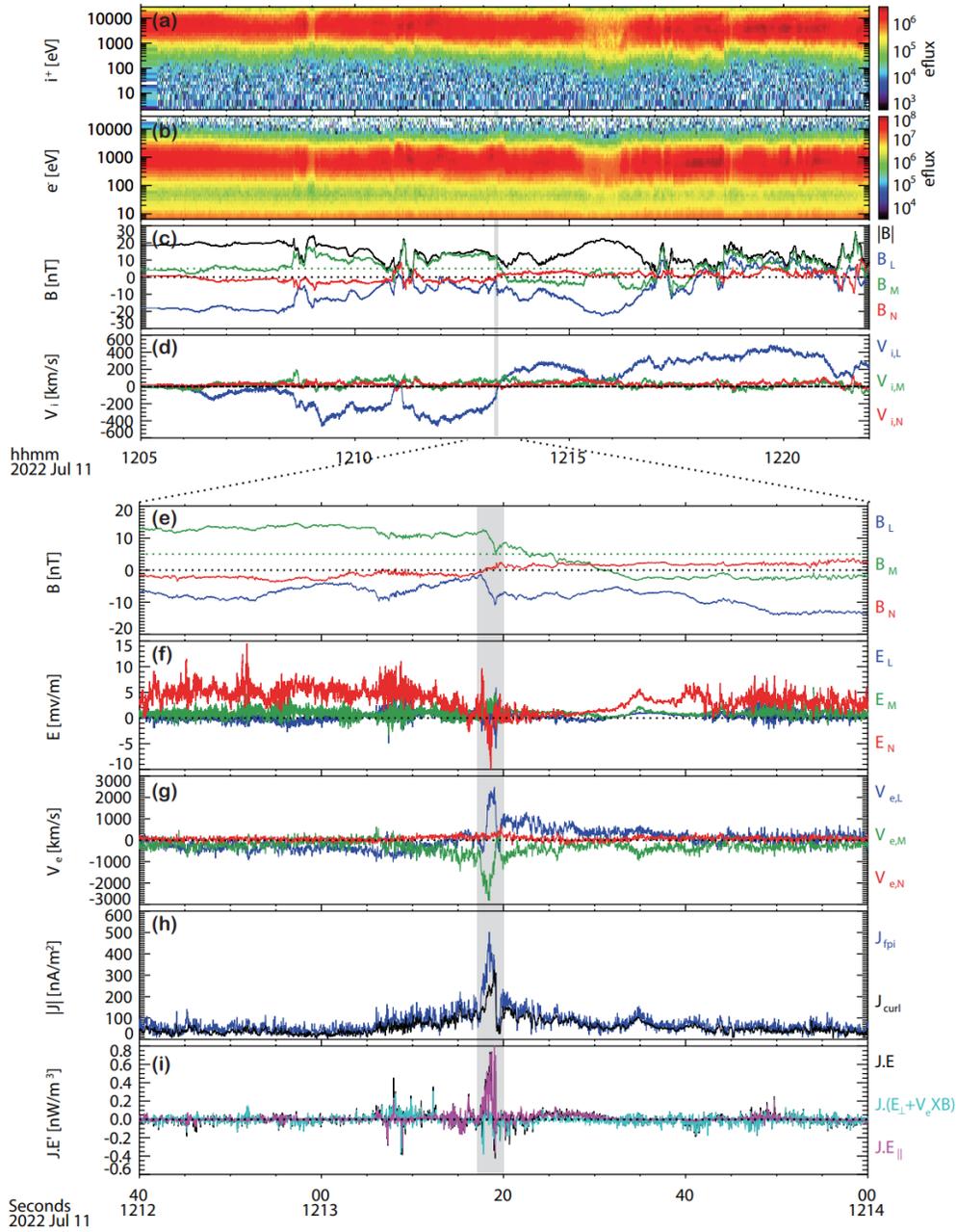

**Figure 1. An overview of the reconnecting current sheet observed by MMS1.** (**a**) and (**b**) Ion and electron differential energy flux spectrum. (**c**) Magnetic field. (**d**) Ion bulk flow velocity. (e)-(i) show the zoomed-in view around the X-line. (**e**) Magnetic field. (**f**) Electric field. (**g**) Electron bulk flow velocity. (**h**) The magnitude of current density. The blue trace is based on the plasma

measurements and black traces are obtained from the curlometer method. **(i)** Energy dissipation rate $J \cdot (E + V_e \times B)$.

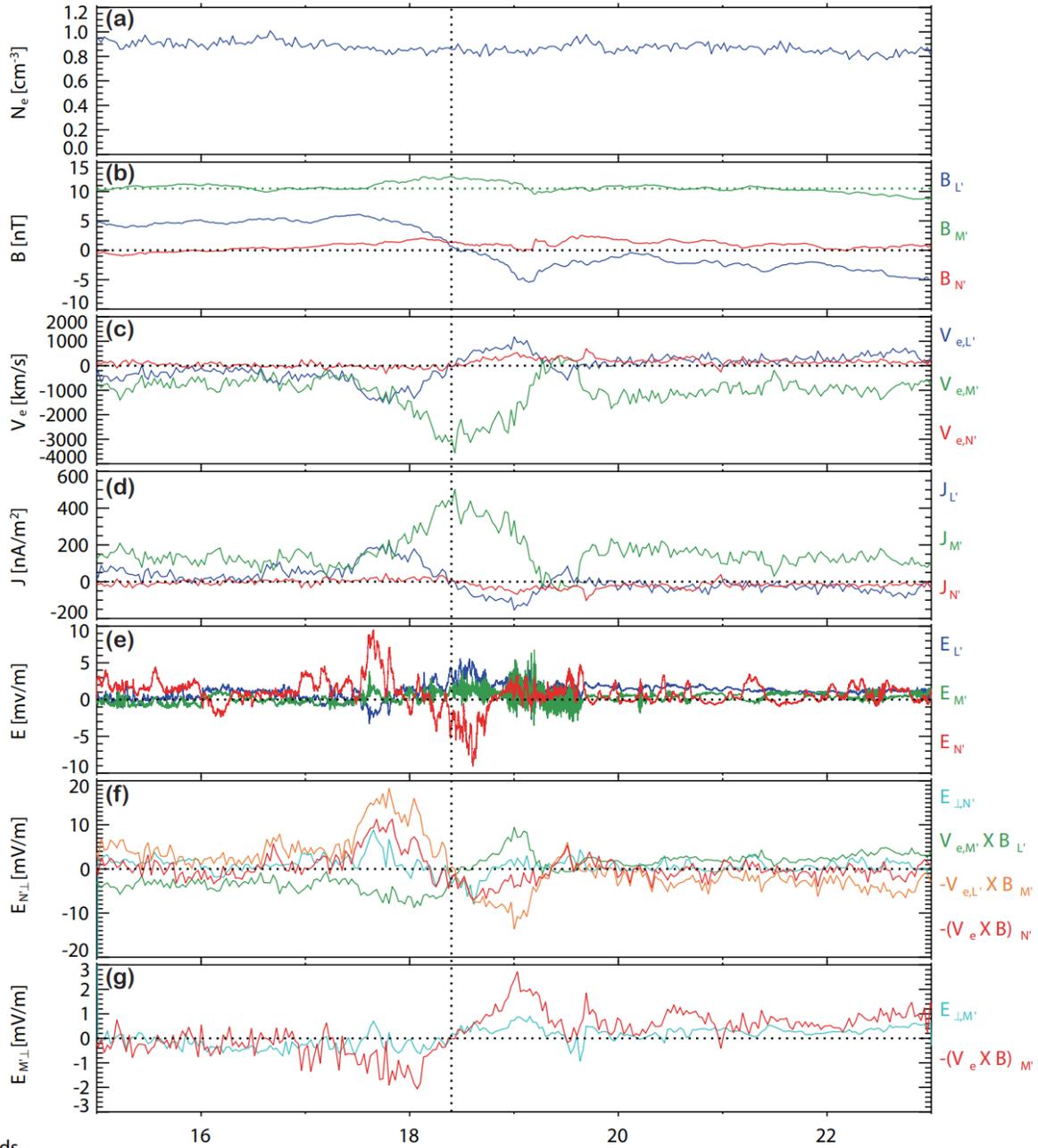

**Figure 2. A zoomed-in view of the EDR LMN' coordinate. (a)** Electron number density. **(b)** Magnetic field. **(c)** Electron bulk flow velocity. **(d)** The three components of current density. **(e)** Electric field. **(f)-(g)** Perpendicular electric field $\mathbf{E}_\perp$, $-(\mathbf{V_e} \times \mathbf{B})$, and $-(\mathbf{V_i} \times \mathbf{B})$.

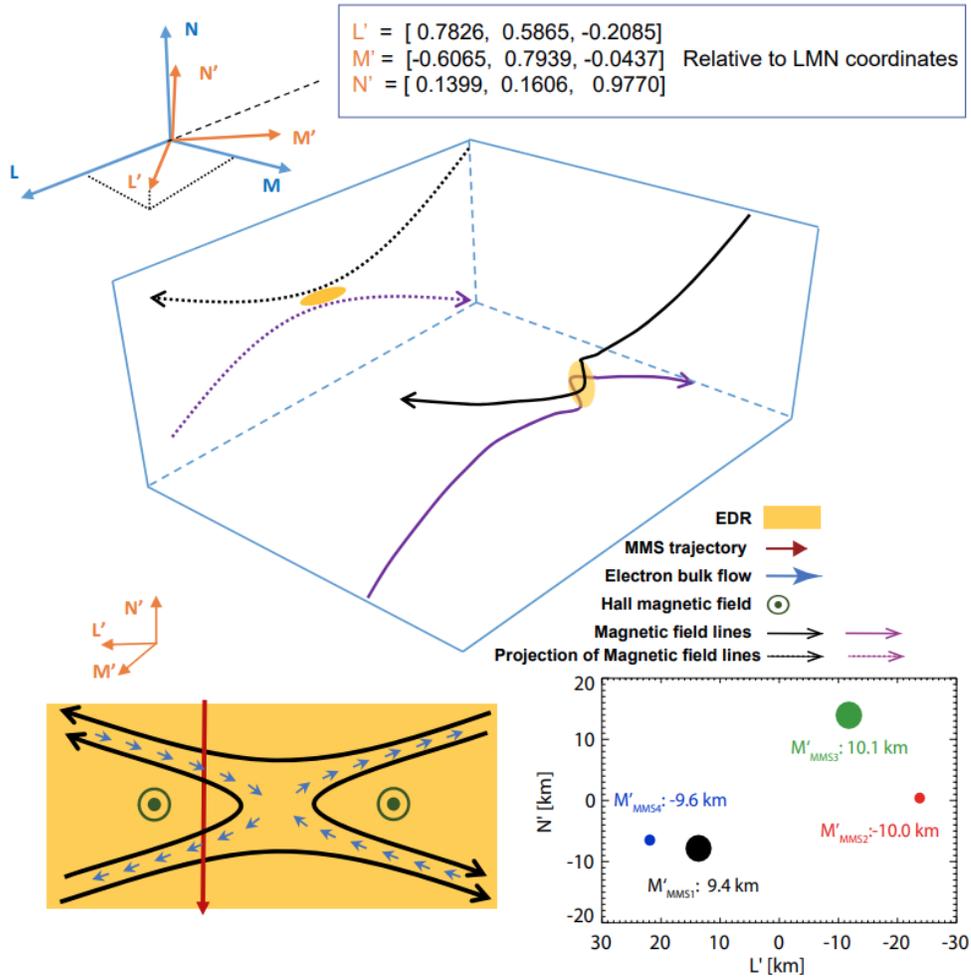

**Figure 3. A schematic illustration of the knotted EDR and the relative location of the MMS spacecraft.**

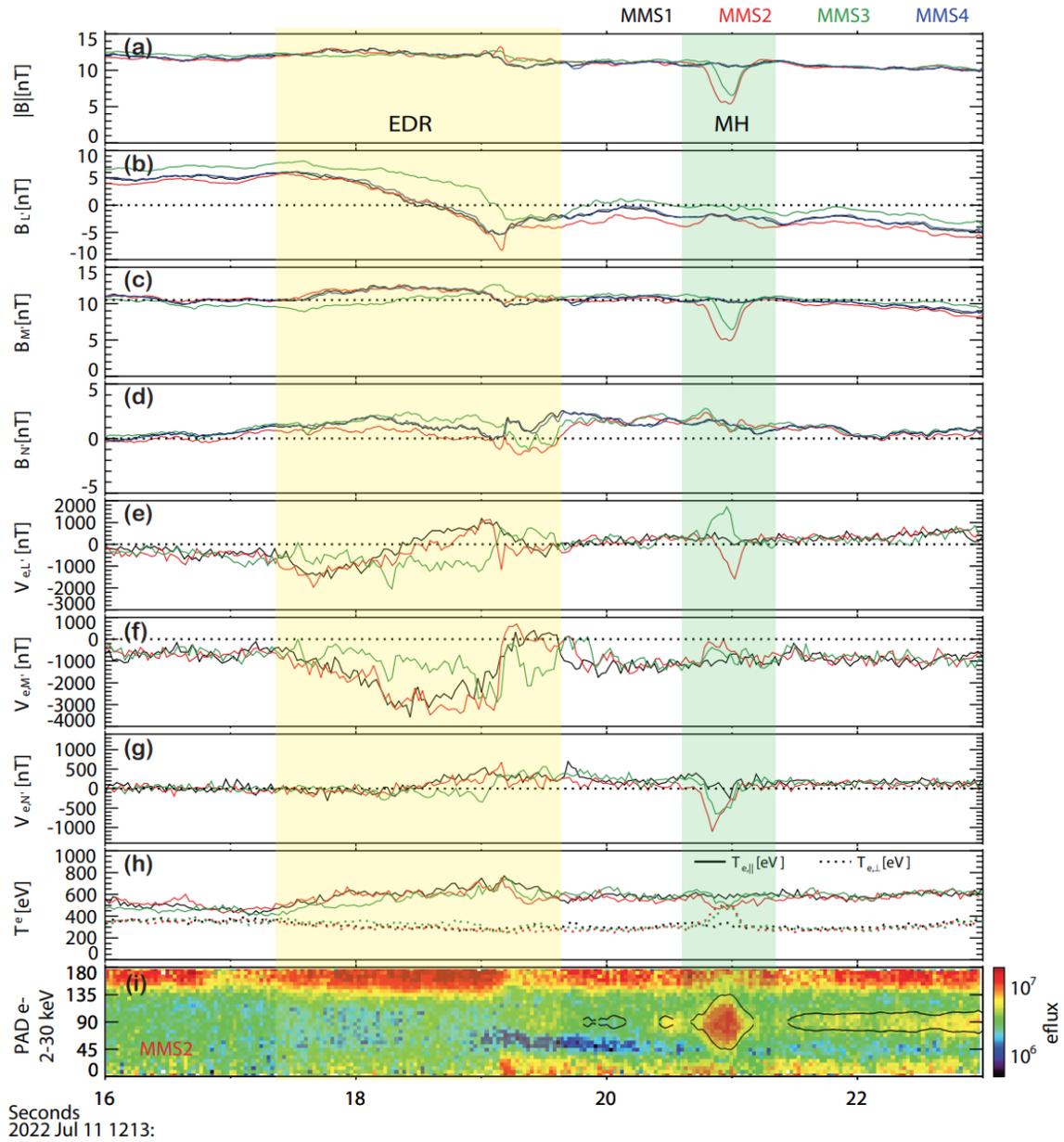

**Figure 4. Multiple spacecraft observations of the knotted EDR. (a)-(d)** Magnitude and three components of the magnetic field. **(e)-(g)** Three components of the electron bulk flow velocity. **(h)** Electron temperature in parallel (solid line) and perpendicular (dashed lines) direction. **(i)** Electron pitch angle distribution in the energy range of 2-30 keV observed by MMS2. The black lines represent a local loss cone angle.